\documentclass[12pt]{spieman}  
\usepackage{amsmath,amsfonts,amssymb}
\usepackage{graphicx}
\usepackage{setspace}
\usepackage{tocloft}
\usepackage{comment}
\usepackage{color}
\usepackage{soul}
\usepackage{url}
\usepackage{physics}

\title{Noise reduction in suspension control with photon-pressure actuator for CHRONOS gravitational wave detector}

\author[a,b,c\dag]{Daiki Tanabe}
\author[b,a,c,d\ddag]{Yuki Inoue}
\author[e,b,a]{Mario Juvenal S. Onglao III}

\affil[a]{Center for High Energy and High Field Physics, National Central University, Taoyuan 32001, Taiwan}
\affil[b]{Physics Department, National Central University, Taoyuan 32001, Taiwan}
\affil[c]{Institute of Particle and Nuclear Studies (IPNS), High Energy Accelerator Research Organization (KEK), Tsukuba, Ibaraki 305-0801, Japan}
\affil[d]{Institute of Physics, Academia Sinica, Nangang, Taipei, 015011, Taiwan}
\affil[e]{National Institute of Physics, University of the Philippines Diliman, Philippines}

\cftpagenumbersoff{figure}
\cftpagenumbersoff{table} 

\begin{document} 
\maketitle

\begin{abstract}
Improving sub-Hz sensitivity of gravitational wave (GW) detectors is important to detect heavier binary black hole mergers and study phenomena in stronger gravity fields. Torsion-bar-based GW detectors have been projected to focus on low-frequency GW. Among noise sources of GW detector, actuation noise induced by vibration of force sources and fluctuation of environmental magnetic fields is one that increases in low frequency. In this study, we propose photon-pressure actuator as a solution to isolate an actuator from seismic and magnetic noise. It can also be used as a photon calibrator. We designed an optical layout of the photon-pressure actuator having four beams independently controlled and applied it to CHRONOS experiment. Based on a realistic power control system, we estimated its maximum torque amplitude around yaw rotation axis as $1.0\times 10^{-8}$~N$\cdot$m and actuation efficiency as $6.6\times 10^{-13}$~rad/V, which are sufficiently large for controlling the CHRONOS torsion bar. The actuation noise was estimated as $5.3\times 10^{-19}$~rad~${\rm Hz}^{-1/2}$ at 1~Hz, lower than the target sensitivity of CHRONOS. Assuming its usage as a photon calibrator, the estimated systematic error was 1.14\%.
\end{abstract}


\keywords{gravitational wave, torsion bar, photon pressure, actuator, intensity noise}

{\noindent \footnotesize\textbf{\dag}Daiki Tanabe,  \linkable{tana2431.ts@gmail.com} }
{\noindent \footnotesize\textbf{\ddag}Corresponding author: Yuki Inoue,  \linkable{iyuki@ncu.edu.tw} }

\begin{spacing}{1}

\section{Introduction}
\label{sect:intro_section}

Sustained improvements in the sensitivity of gravitational-wave (GW) detectors over the past decades have brought a transition of the research field from the initial stage of discovery to an era of precision measurements. The first season of the fourth joint observing run (O4) of LIGO, Virgo, and KAGRA, which concluded in 2024, reported 128 candidate GW events~\cite{gwtc4}. Statistical analyses of this growing event catalog have enabled to draw the mass distribution of compact binaries, yielding results that contribute to cosmology, for example by testing new methods toward determining the Hubble constant~\cite{gwtc4-population,gwtc4-hubble}.

Further progress in cosmology by GW requires extending the detectable mass range of GW sources, in addition to increasing the number of detected GW events. Black holes with masses in the range of $\mathcal{O}(10^2)$--$\mathcal{O}(10^5)$M$_\odot$, referred to as intermediate-mass black holes (IMBHs), are expected as a potential missing link between stellar-mass black holes and supermassive black holes with masses above $\mathcal{O}(10^5)$M$_\odot$ in galactic nuclei. Only a small number of IMBH binaries have been detected as GW sources to date. Furthermore, IMBH binaries with component masses above $\mathcal{O}(10^3)$M$_\odot$ exert GW at frequencies typically below the sub-hertz band. It implies that their detection requires improvement of low-frequency sensitivity in GW detectors.

Several interferometer configurations specifically targeting low-frequency sensitivity have been proposed based on torsion-bar test masses, including TOBA, TorPeDO, and CHRONOS~\cite{toba,Torpedo,CHRONOS}. These experiments sense the rotational, instead of translational, motion of bar-shaped end-test masses (ETMs). Rotational modes of a torsion bar can be designed to have substantially reduced resonance frequencies, and therefore can be sensitive to GW signals below 1~Hz. CHRONOS further aims to enhance sub-hertz sensitivity by combining a torsion-bar design with a speed-meter topology and cryogenic mirrors, with the goal of detecting IMBH binaries.

Seismic noise is one of the most significant among the noise sources that limit low-frequency sensitivity. It couples not only directly to the ETM but also indirectly through the actuator used to control the ETM position and to maintain interferometer lock. In conventional GW detectors, a recoil mass is separately suspended near the test mass to apply forces via electromagnetic or electrostatic actuators~\cite{kagra_actuation,kagra_coilmagnet,ligo_electrostatic_parametric}. This configuration, however, allows excess actuation noise to arise from the vibration of the recoil mass. The electromagnetic actuator additionally couples with the environmental magnetic fields to induce low-frequency noise. A coil-coil actuator deployed in TOBA is free from magnetic noise but still affected by mechanical vibration~\cite{toba_oshima}.

We propose a photon-pressure actuator in this work to mitigate the influence of both seismic and magnetic noise related to the ETM control. The momentum of photons is insensitive to mechanical vibrations of the light source, does not couple to magnetic fields, and is not attenuated by distances. As a consequence, it has the potential to simplify suspension structures by eliminating the recoil mass.

The idea of actuating an ETM by photon pressure has been studied and deployed extensively in the context of Photon Calibrators (PCals)~\cite{ligo_pcal,virgo_pcal,kagra_pcal}. The main feature of PCal is controlling power and position of the laser beams to inject them to the test mass. The force and torque applied by the beam has been modeled by a simple formalization. The latest PCal designed by KAGRA controls two beams independently to measure rotational response of the mirror which is one of the major sources of calibration error~\cite{kagra_pcal,dripta_2023sep}. Based on the techniques from PCal, we propose a system that controls four beams independently for controlling the degrees of freedom of translation, yaw, and pitch. 

In addition to mitigating the low frequency actuation noise, the photon-pressure actuator can be integrated with PCal because they work in the same principle. When it is designed as a two-in-one system, it realizes further simplified hardware and operation. Particularly, its accurately monitored force simplifies a model of actuation transfer function. 

Fundamental challenge for the photon-pressure actuator is to ensure sufficient force to actuate the ETM while producing noise lower than the target sensitivity of detector. Force generated by a PCal laser beam, typically at a few watt power, is smaller than that of an electromagnetic actuator. We evaluated actuation range, noise, and systematic error when used as PCal based on parameters of realistic hardware components. We chose CHRONOS as a platform of design because of its significant demand to low noise at the low frequencies.

To our knowledge, this is the first dedicated study of photon-pressure actuator as a means to ETM control without coupling to ground motion and magnetic fields in a GW detector. Section~\ref{sect:actuator} describes the operating principle of the photon-pressure actuator with an analytical model. Section~\ref{sect:design} presents an optical design that can be implemented in a realistic CHRONOS configuration. Section~\ref{sect:range} calculates the maximum force and ETM rotation angle induced by this force. Section~\ref{sect:uncertainty} evaluates actuator noise and systematic uncertainties.

\section{Actuation principle of photon-pressure actuator}
\label{sect:actuator}

\begin{figure}[tp]
\begin{center}
\begin{tabular}{c}
\includegraphics[width=8.0cm]{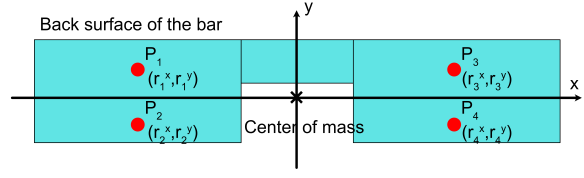}
\end{tabular}
\end{center}
\caption[]
{ \label{fig:4mirrors}
Conceptual positions and indices of the beams injected from the photon-pressure actuator onto the back surface of CHRONOS torsion bar.
}
\end{figure}

The basic idea of the photon pressure actuator is to inject power-controlled laser beams to the test mass to actuate it by the photon pressure. In order to realize actuation of yaw, pitch, and translational degrees of freedom without interfering with the beam path for the main interferometer, we inject four beams to the back surface of our ETM. These injection points should be approximately-symmetrically arranged around the center of mass of the bar as shown in Fig.~\ref{fig:4mirrors}, although we should slightly adjust the positions to align them to the nodal lines of the internal resonant mode of the bar. We label the four injection points by index ${\rm i}=1,2,3,4$ such that the beam power at each point is denoted by $P_{\rm i}$. We define the x-y plane by the back surface of the bar and z axis by the translational direction, spanned by unit vectors $\vb{e}_x$, $\vb{e}_y$, $\vb{e}_z$. The position vector from the center of mass is written as $\vb{r}_{\rm i}=(r_{\rm i}^x, r_{\rm i}^y,0)$. 

We parameterize the force applied by each beam by beam power $P_{\rm i}$, azimuthal incident angle $\theta_{\rm i}$, and inclination incident angle $\phi_{\rm i}$. Using these parameters, the force vector of the beam can be written as
\begin{equation}
\label{eq:pcal_force}
\vb{F}_{\rm i}=\left(\frac{2P_{\rm i}(t)\cos\phi_{\rm i}\sin\theta_{\rm i}}{c},\frac{2P_{\rm i}(t)\sin\phi_{\rm i}}{c},\frac{2P_{\rm i}(t)\cos\phi_{\rm i}\cos\theta_{\rm i}}{c} \right),    
\end{equation}
where $c$ is the speed of light. Therefore, the net force in the translational direction and torques in the yaw and pitch rotation are
\begin{equation}
\label{eq:net_force}
F_{\rm T}(t)=\sum_{\rm i}^4 \vb{F}_{\rm i}\cdot \vb{e}_z,
\end{equation}
\begin{equation}
\label{eq:net_torque_yaw}
T_{\rm Y}(t)=\sum_{\rm i}^4 \left( \vb{r}_{\rm i}\times\vb{F}_{\rm i} \right)\cdot \vb{e}_y,
\end{equation}
\begin{equation}
\label{eq:net_torque_pitch}
T_{\rm P}(t)=\sum_{\rm i}^4 \left( \vb{r}_{\rm i}\times\vb{F}_{\rm i} \right)\cdot \vb{e}_x.
\end{equation}

The equation of motion of translational displacement and rotations can be written as
\begin{equation}
\label{eq:trans_t}
M\ddot{z}(t)+\gamma \dot{z}(t)+kz(t)=F_{\rm T}(t),
\end{equation}
\begin{equation}
\label{eq:yaw_t}
I_{\rm Y}\ddot{\varphi}_{\rm Y}(t)+\Gamma_{\rm Y} \dot{\varphi}_{\rm Y}(t)+K_{\rm Y}\varphi_{\rm Y}(t)=T_{\rm Y}(t),
\end{equation}
\begin{equation}
\label{eq:pitch_t}
I_{\rm P}\ddot{\varphi}_{\rm P}(t)+\Gamma_{\rm P} \dot{\varphi}_{\rm P}(t)+K_{\rm P}\varphi_{\rm P}(t)=T_{\rm P}(t),
\end{equation}
where $M$ is mass of the bar, $I_{\rm Y}$ and $I_{\rm P}$ are moment of inertia around each rotation axis, $\gamma$ and $\Gamma$ are damping coefficients, $k$ and $K$ are restoring coefficients originated to gravity and torsion spring coefficient. 

The Eqs.~(\ref{eq:trans_t})(\ref{eq:yaw_t})(\ref{eq:pitch_t}) can be converted into the frequency domain by Laplace transform as
\begin{equation}
\label{eq:trans_f}
-M\omega^2 z(\omega)+i\omega\gamma z(\omega)+kz(\omega)=F_{\rm T}(\omega),  
\end{equation}
\begin{equation}
\label{eq:yaw_f}
-I_{\rm Y}\omega^2\varphi_{\rm Y}(\omega)+i\omega\Gamma_{\rm Y} \varphi_{\rm Y}(\omega)+K_{\rm Y}\varphi_{\rm Y}(\omega)=T_{\rm Y}(\omega),
\end{equation}
\begin{equation}
\label{eq:pitch_f}
-I_{\rm P}\omega^2\varphi_{\rm P}(\omega)+i\omega\Gamma_{\rm P} \varphi_{\rm P}(\omega)+K_{\rm P}\varphi_{\rm P}(\omega)=T_{\rm P}(\omega),
\end{equation}
where $\omega\equiv 2\pi f$ for the frequency $f$. Therefore, the displacement and rotation angles in the frequency domain are
\begin{equation}
\label{eq:trans}
\begin{array}{ll}
z(\omega)&=\frac{F_{\rm T}(\omega)}{-M\omega^2+i\omega\gamma+k}=\frac{F_{\rm T}(\omega)}{-M\left( \omega^2-i\frac{\Omega_{\rm T} \omega}{Q_{\rm T}}-\Omega_{\rm T}^2 \right)} \\
&\simeq -\frac{F_{\rm T}(\omega)}{M\omega^2},
\end{array}
\end{equation}
\begin{equation}
\label{eq:yaw}
\begin{array}{ll}
\varphi_{\rm Y}(\omega)&=\frac{T_{\rm Y}(\omega)}{-I_{\rm Y}\omega^2+i\omega\Gamma_{\rm Y}+K_{\rm Y}}=\frac{T_{\rm Y}(\omega)}{-I_{\rm Y}\left( \omega^2-i\frac{\Omega_{\rm Y} \omega}{Q_{\rm Y}}-\Omega_{\rm Y}^2 \right)} \\
&\simeq -\frac{T_{\rm Y}(\omega)}{I_{\rm Y}\omega^2},
\end{array}
\end{equation}
\begin{equation}
\label{eq:pitch}
\begin{array}{ll}
\varphi_{\rm P}(\omega)&=\frac{T_{\rm P}(\omega)}{-I_{\rm P}\omega^2+i\omega\Gamma_{\rm P}+K_{\rm P}}=\frac{T_{\rm P}(\omega)}{-I_{\rm P}\left( \omega^2-i\frac{\Omega_{\rm P} \omega}{Q_{\rm P}}-\Omega_{\rm P}^2 \right)} \\
&\simeq -\frac{T_{\rm P}(\omega)}{I_{\rm P}\omega^2},
\end{array}
\end{equation}
where
\begin{equation}
\Omega_{\rm T}\equiv \sqrt{\frac{k}{M}}, \ \Omega_{\rm Y}\equiv \sqrt{\frac{K_{\rm Y}}{I_{\rm Y}}}, \ \Omega_{\rm P}\equiv \sqrt{\frac{K_{\rm P}}{I_{\rm P}}},    
\end{equation}
\begin{equation}
Q_{\rm T}\equiv \frac{\Omega_{\rm T}M}{\gamma}, \ Q_{\rm Y}\equiv \frac{\Omega_{\rm Y}I_{\rm Y}}{\Gamma_{\rm Y}}, \ Q_{\rm P}\equiv \frac{\Omega_{\rm P}I_{\rm P}}{\Gamma_{\rm P}}.
\end{equation}
The $\Omega$ represent the resonant frequencies, and the $Q$ represents the quality factors of each degree of freedom (DoF). The resonant frequencies of the CHRONOS torsion bar are less than 1~Hz, and the quality factor of sapphire is typically at the order of $10^8$. Therefore, we neglected the $\Omega/Q$ and $\Omega^2$ terms in Eqs.~(\ref{eq:trans})(\ref{eq:yaw})(\ref{eq:pitch}).

\section{Optical configuration}
\label{sect:design}

In order to realize the four-beams injection with powers and positions that are independently controlled, we propose optics based on acousto-optic modulators (AOMs) coupled with optical follower servos (OFSs) and digital-to-analog converters (DACs) for control. This configuration has been demonstrated in PCal studies in LIGO and KAGRA~\cite{ligo_pcal,kagra_pcal}. The OFS generates analog feedback signal to the AOM and also allows external input for changing the offset voltage.

\begin{figure}[tp]
\begin{center}
\begin{tabular}{c}
\includegraphics[width=15.0cm]{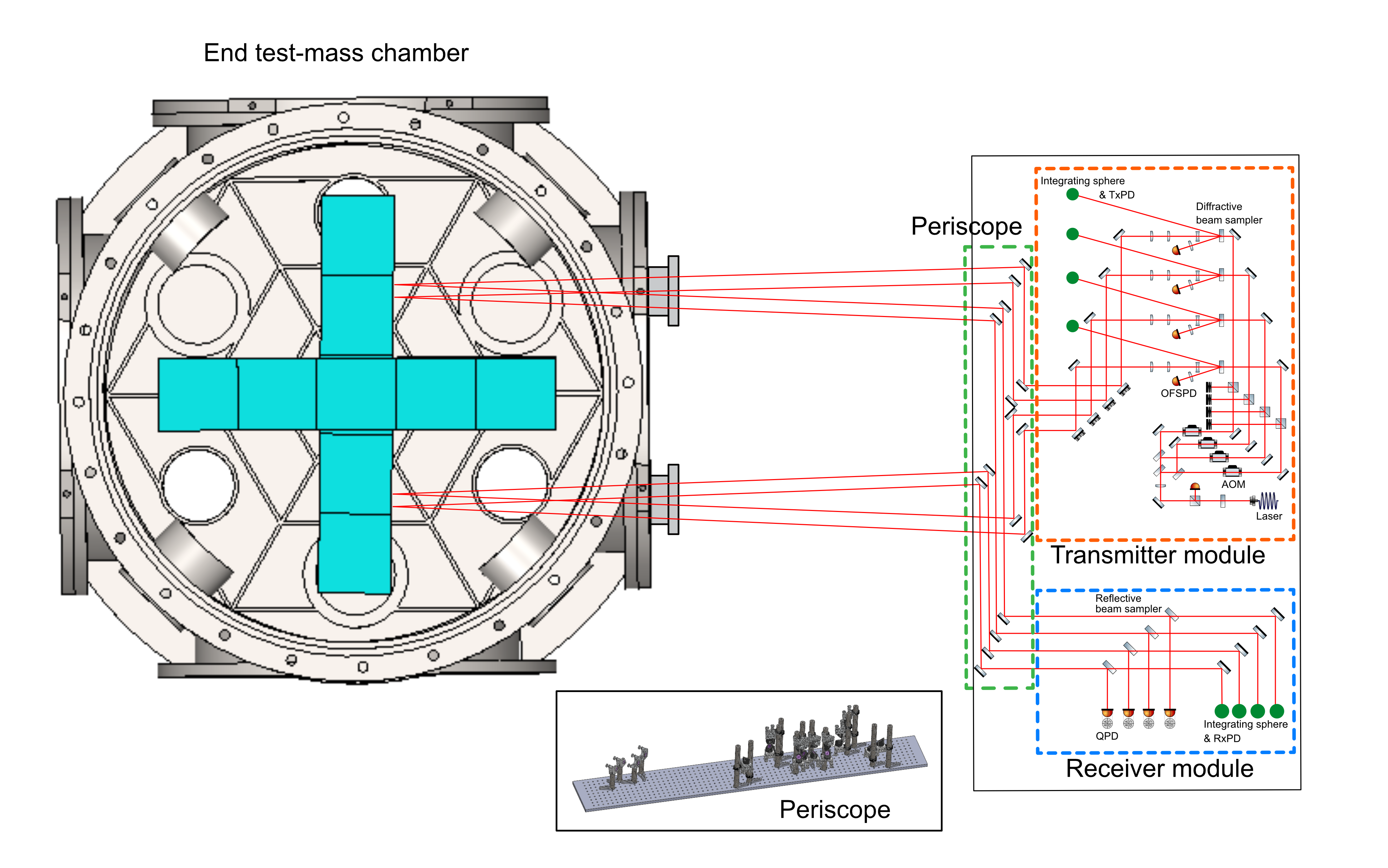}
\end{tabular}
\end{center}
\caption[]
{ \label{fig:overall}
Conceptual optical layout of the photon-pressure actuator with the ETM chamber of CHRONOS. Optical setup for only one bar is shown. See Fig.~\ref{fig:transmitter} and Fig.~\ref{fig:receiver} for the detailed views of the transmitter and receiver modules.
}
\end{figure}

Figure~\ref{fig:overall} shows a conceptual optical layout of the four-beams photon pressure actuator. It consists of the transmitter module, periscope, and receiver module. The transmitter module controls the beam power, while the receiver module monitors the reflected power and the beam positions. Beam orientations are adjusted at the periscope to hit the designated positions on the ETM.

\begin{figure}[tp]
\begin{center}
\begin{tabular}{c}
\includegraphics[width=8.0cm]{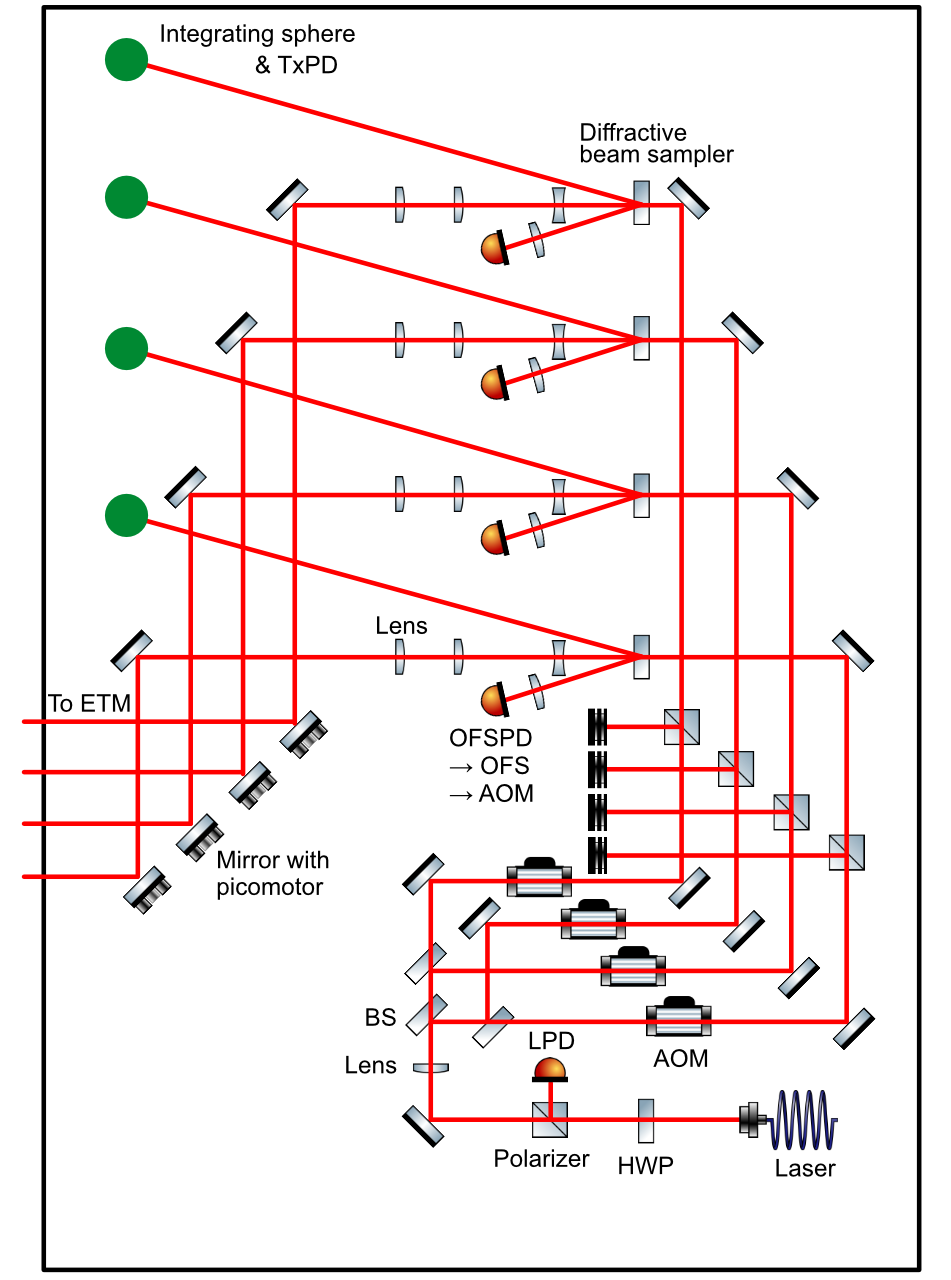}
\end{tabular}
\end{center}
\caption[]
{ \label{fig:transmitter}
Conceptual optical layout of the transmitter module for photon pressure actuator.
}
\end{figure}

Figure~\ref{fig:transmitter} shows the schematic diagram of the transmitter module. Powers and positions of the four beams should be independently controlled. We measure each beam power by a photodetector labeled as OFSPD. Each of four OFSPDs is connected to an OFS to feedback the intensity signal to AOM. Out of the feedback loop, an independent photodetector labeled as TxPD monitors each power. Each TxPD is coupled with an integrating sphere to collect entire power of the beam. Beams for power monitoring are sampled from the transmitted beams by diffractive beam samplers. The mirrors at the final part of the transmitter module have picomotors to control orientations of the transmitted beams.

To control the actuation power, DACs provide modulation signal from outside of the optical system to the OFS. The DAC also supplies an offset voltage to OFS to determine the operation point of AOM.

\begin{figure}[tp]
\begin{center}
\begin{tabular}{c}
\includegraphics[width=8.0cm]{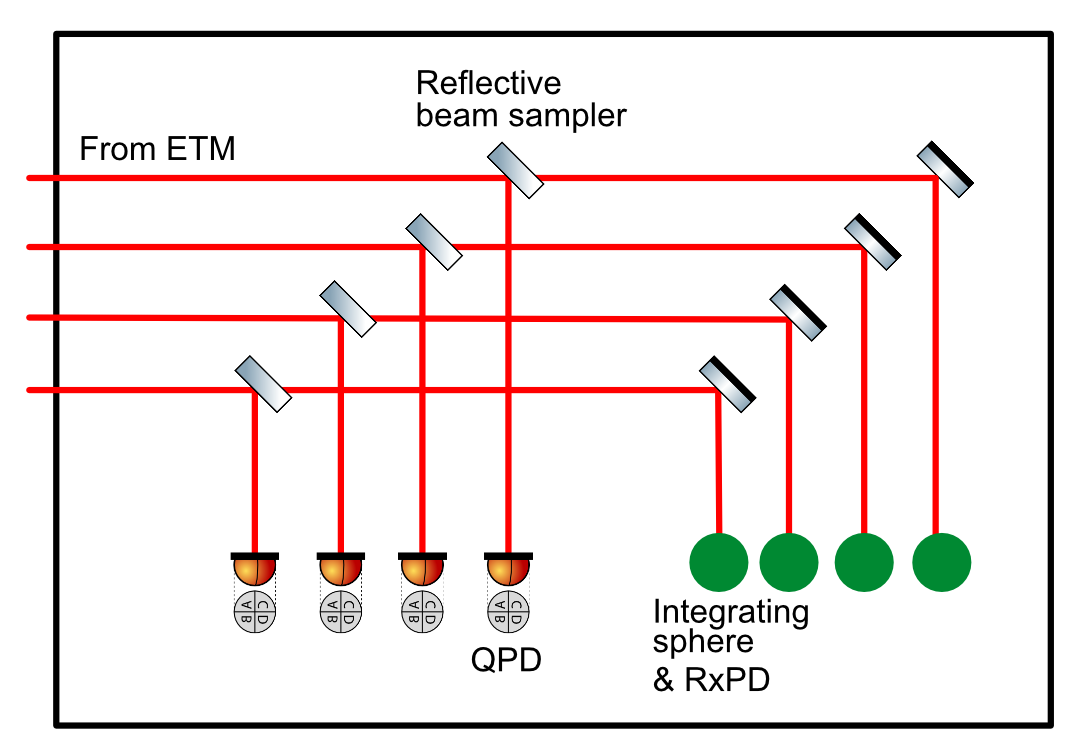}
\end{tabular}
\end{center}
\caption[]
{ \label{fig:receiver}
Conceptual optical layout of the receiver module for the photon pressure actuator.}
\end{figure}

The beams reflected by the ETM are received by the receiver module. Figure~\ref{fig:receiver} shows the schematic diagram of the receiver module. In the receiver module, the total power of the four beams is monitored by a photodetector labeled as RxPD, coupled with an integrating sphere. A small fraction of the beam is sampled by beam samplers and led to quadrant photodetectors for monitoring the beam positions. 

\section{Actuation range}
\label{sect:range}

\subsection{Maximum actuation force}
\label{sect:actuation_force}

Here we evaluate the actuation range of the photon-pressure actuator with the realistic parameter values for CHRONOS as summarized in Table~\ref{tab:parameters}. Incident angle on the torsion bar is constrained by the chamber size and the window size. We inject the beams through the window attached to the CF100 vacuum flange and let the beam reflected back through the same window. Considering the effective diameter of the window is approximately 0.08~m and it is 0.8~m away from the surface of the torsion bar, the incident angle is 2.9$^\circ$ at most. We assume 2.9$^\circ$ for the azimuthal incident angle $\theta_{\rm i}$ and 0$^\circ$ for the inclination $\phi_{\rm i}$. The mass of the torsion bar is 171~kg, and the moment of inertia is 19.9~kg$\cdot$m$^2$ in yaw and 1.3~kg$\cdot$m$^2$ in pitch~\cite{CHRONOS}.

\begin{table}[tbp]
\begin{center}\footnotesize
\begin{tabular}{c|c|c}
\hline
Parameter & Symbol & Value (Unit) \\
\hline\hline
Max amplitude of beam power & $P_{\rm i}$ & 1.25 (W) \\
Azimuthal incident angle & $\theta_{\rm i}$ & $\le$2.9 ($^\circ$) \\
Inclination incident angle & $\phi_{\rm i}$ & $0$ ($^\circ$) \\
Upper beam position (i=1,3) & $(r_{\rm i}^x,r_{\rm i}^y)$ & ($\mp$0.306,0.053) (m,m) \\
Lower beam position (i=2,4) & $(r_{\rm i}^x,r_{\rm i}^y)$ & ($\mp$0.317,-0.021) (m,m) \\
Mass of the bar & $M$ & 171 (kg) \\
Moment of inertia of the bar (yaw) & $I_{\rm Y}$ & 19.9 (kg$\cdot$m$^2$) \\
Moment of inertia of the bar (pitch) & $I_{\rm P}$ & 1.3 (kg$\cdot$m$^2$) \\
\hline
\end{tabular}
\end{center}
\caption{Parameters related to the photon-pressure actuator for CHRONOS.}
\label{tab:parameters}
\end{table}

The injection positions should be on the nodal line of the internal resonant modes of the bar because the excitation of these modes can cause spurious rotation signal. The CHRONOS torsion bar has resonant modes at 667~Hz and 1237~Hz, where the first mode is the vertical bending mode and the second is the horizontal bending mode~\cite{tanabe_chronos_intensity}. The second mode can couple to the rotation signal, particularly when the main interferometer beams are at asymmetric positions due to unexpected beam offset.

\begin{figure}[tp]
\begin{center}
\begin{tabular}{c}
\includegraphics[width=8.0cm]{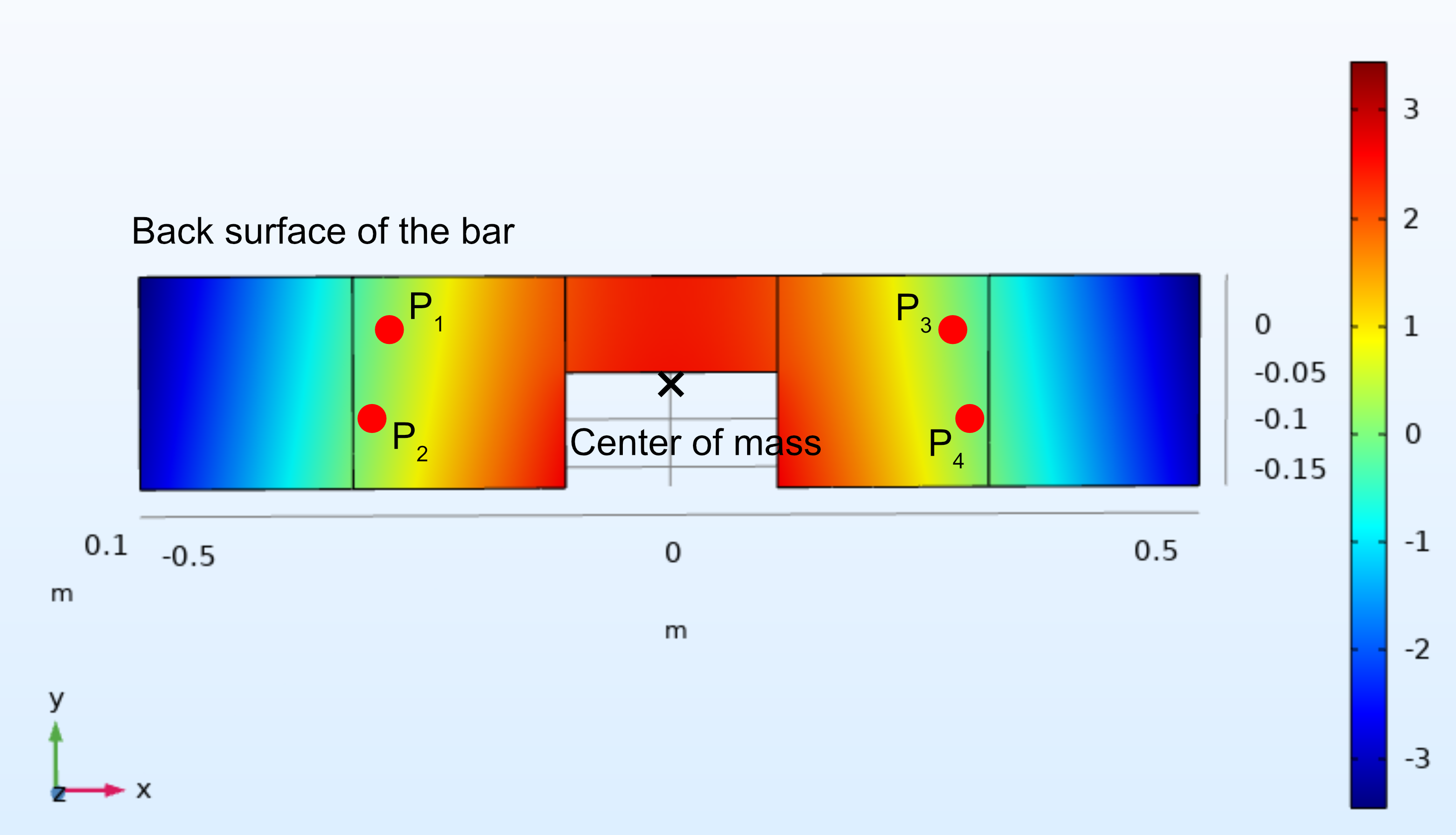}
\end{tabular}
\end{center}
\caption[]
{ \label{fig:injection_node}
Injection positions of the CHRONOS photon-pressure actuator beams and the mode shape of the horizontal bending mode of the CHRONOS torsion bar at 1238~Hz simulated by COMSOL. Unit of the color legend is arbitrary.
}
\end{figure}

The beams are injected to 1-inch mirrors attached on the back surface of ETM. The mirrors should be away from the seams of the sapphire blocks to ensure optical uniformity. To realize the maximum distance from the center of mass under these constraints, we determine the injection positions as $(r_1^x,r_1^y)=(-0.306,0.051)$, $(r_2^x,r_2^y)=(-0.321,-0.046)$, $(r_3^x,r_3^y)=(0.306,0.051)$, $(r_4^x,r_4^y)=(0.321,-0.046)$ in the unit of meter. Figure~\ref{fig:injection_node} shows the shape of the 1238~Hz mode and the injection positions coinciding with the nodal line of this mode. The mode frequency and shape are simulated by COMSOL~\cite{comsol}.

We deploy a FC-1550-10W provided by CNI, a fiber-coupled 10~W constant-wave laser with a wavelength of 1550~nm, as a laser source~\cite{Laser_FC-1550-10W}. It is split into four beams of 2.5~W power by beam splitters and injected to AOM. We modulate it with the excitation signal $V_{\rm exc}$ sent from DAC to OFS. As calculated in detail in Sec.~\ref{sect:appendix_parameters}, the actuation power range is $\pm$1.25~W centered at 1.25~W.

We assume that the beams 3 and 4 are modulated in the opposite phase of the beams 1 and 2 when it is intended to actuate in the yaw degree of freedom. Despite the asymmetry of the beam positions in the vertical direction, we don't adjust the power to balance torque because it only makes offset of the center of the pitch rotation. 

Substituting these values into Eq.~(\ref{eq:pcal_force}), we derive $8.3\times 10^{-9}$~N as the maximum force amplitude that each beam can apply. As summarized in Table~\ref{tab:efficiency}, the maximum total force amplitude in the translational direction is $3.3\times 10^{-8}$~N, while the maximum total torque amplitude is $1.0\times 10^{-8}$~N$\cdot$m in yaw and $1.2\times 10^{-9}$~N$\cdot$m in pitch. The maximum force and torque are twice as these amplitudes.

\begin{table}[tbp]
\begin{center}\scriptsize
\begin{tabular}{c|c|c|c}
\hline
DoF & \begin{tabular}{c} Max force or\\torque amplitude\\(Unit) \end{tabular} & \begin{tabular}{c} Max actuation\\efficiency (Unit) \end{tabular} & \begin{tabular}{c} Noise (Unit) \end{tabular} \\
\hline\hline
Trans. & 3.3$\times$10$^{-8}$ (N) & 2.5$\times 10^{-13}$ (m/V) & 3.9$\times 10^{-19}$ (m/$\sqrt{\rm Hz}$) \\
Yaw & 1.0$\times$10$^{-8}$ (N$\cdot$m) & 6.6$\times$10$^{-13}$ (rad/V) & 5.3$\times 10^{-19}$ (rad/$\sqrt{\rm Hz}$) \\
Pitch & 1.2$\times$10$^{-9}$ (N$\cdot$m) & 1.2$\times$10$^{-12}$ (rad/V) & 1.1$\times 10^{-18}$ (rad/$\sqrt{\rm Hz}$) \\
\end{tabular}
\end{center}
\caption{Actuation force, efficiency, and noise in each degree of freedom of CHRONOS. Torque is shown for yaw and pitch rotation instead of force. The actuation efficiency and noise was evaluated at 1~Hz.}
\label{tab:efficiency}
\end{table}

\subsection{Actuation efficiency}
\label{sect:actuation_efficiency}

We can calculate the maximum displacement and angle that the photon-pressure actuator can generate by substituting the maximum force derived in Sec.~\ref{sect:actuation_force} into Eqs.~(\ref{eq:trans})(\ref{eq:yaw})(\ref{eq:pitch}). The maximum amplitude of translational displacement is 4.9$\times 10^{-12}$~m while that of rotational angle is 1.3$\times 10^{-11}$~rad in yaw and 2.4$\times 10^{-11}$~rad in pitch, respectively.

We control the beam power by adjusting the input voltage of AOM through the OFS offset and excitation voltage, which are provided by DAC. In contrast to electromagnetic or electrostatic actuators, a voltage supplier does not fundamentally determine the maximum force that the photon-pressure actuator can apply, although we can still define it for comparison as the maximum motion per DAC voltage.

We assigned the actuation power range of $\pm$1.25~W to the DAC output of $\pm$20~V in Sec.~\ref{sect:appendix_parameters}. It leads to the force as 8.3$\times 10^{-10}$~N per beam per unit DAC voltage. Substituting them to Eqs.~(\ref{eq:trans})(\ref{eq:yaw})(\ref{eq:pitch}), we obtain the actuation efficiency as 2.5$\times 10^{-13}$~m/V in translation, 6.6$\times 10^{-13}$~rad/V in yaw, and 1.2$\times 10^{-12}$~rad/V in pitch. Table~\ref{tab:efficiency} summarizes them.

\section{Uncertainties}
\label{sect:uncertainty}

\subsection{Noise}
\label{sect:noise}

\subsubsection{Power noise}
Every noise in the photon-pressure actuator propagates only through the fluctuation of laser power injected to the ETM. The fundamental noise source is the laser intensity noise that the laser source intrinsically has. In order to suppress this laser intensity noise, we deploy power stabilization based on feedback control using OFS and AOM. However, the feedback circuit induces several other noise into the output laser power; shot noise at PDs and DAC noise. Formalism and evaluation of these noise are described in Appendix~\ref{sect:appendix}.

While the noise induced by each feedback circuit is independent among the four beam paths, the laser intensity noise is in-phase because it comes from the source of all the beams. Therefore, we separately treat the laser intensity noise and other noise induced by feedback to calculate their propagation to the torque. According to the formalism in Eqs.~(\ref{eq:delta_P_laser})(\ref{eq:delta_P_feedback}) and noise evaluation in Sec.~\ref{sect:appendix_noise} with the parameters assumed in Sec.~\ref{sect:appendix_parameters}, the laser intensity noise $\delta P_{\rm laser}$ and $\delta P_{\rm feedback}$ are calculated. They gives the force noise in the direction of ETM surface normal as
\begin{equation}
\delta F_{\rm iz,laser}(\omega)=\frac{2\delta P_{\rm laser}\cos\phi_{\rm i}\cos\theta_{\rm i}}{c},
\end{equation}
\begin{equation}
\delta F_{\rm iz,feedback}(\omega)=\frac{2\delta P_{\rm feedback}\cos\phi_{\rm i}\cos\theta_{\rm i}}{c}.
\end{equation}

When the force noise is converted to torque based on Eq.~(\ref{eq:net_torque_yaw}), the laser intensity noise is canceled out between the left and right of the torsion bar. This canceling effect suppresses the in-phase noise by a factor of beam-position mismatch ratio denoted by $\alpha$~\cite{tanabe_chronos_intensity}. The laser intensity component in the net torque noise is
\begin{equation}
\label{eq:torque_noise_laser}
\delta T_{Y,{\rm laser}}(\omega)=\left\{ r_1^x \delta F_{\rm 1z,laser}(\omega)+r_3^x \delta F_{\rm 3z,laser}(\omega)+r_2^x \delta F_{\rm 2z,laser}(\omega)+r_4^x \delta F_{\rm 4z,laser}(\omega)\right\}\alpha,
\end{equation}
around the yaw rotation axis. Here we assume 1~mm error, which corresponds to $\alpha\sim$0.3\% in horizontal and $\alpha\sim$2.5\% in vertical direction. The canceling effect does not occur in the translation. 

On the other hand, the noise induced by the feedback circuit is summed as quadratures without cancellation.
\begin{equation}
\label{eq:torque_noise_feedback}
\delta T_{Y,{\rm feedback}}(\omega)=\sqrt{\left( r_1^x \delta F_{\rm 1z,laser}(\omega) \right)^2+\left( r_2^x \delta F_{\rm 2z,laser}(\omega) \right)^2+\left( r_3^x \delta F_{\rm 3z,laser}(\omega) \right)^2+\left( r_4^x \delta F_{\rm 4z,laser}(\omega) \right)^2}.
\end{equation}

Hence, the total noise is calculated by
\begin{equation}
\label{eq:torque_noise_total}
\delta T_Y=\sqrt{\delta T_{Y,{\rm laser}}^2(\omega)+\delta T_{Y,{\rm feedback}}^2(\omega)}.
\end{equation}
It generates the yaw angle noise following Eq.~(\ref{eq:yaw}). 

\subsubsection{Magnetic and seismic noise}
Photon pressure actuator does not couple with magnetic environment and the seismic noise. The magnetic field in the circumstance around the ETM and transmitter module does not affect to the laser power because a photon does not have charge. Vibration of the transmitter module does not change the photon pressure due to the constant speed of light.

\subsubsection{Total noise}
Based on Eqs.~(\ref{eq:torque_noise_laser})(\ref{eq:torque_noise_feedback})(\ref{eq:torque_noise_total}) and noise components calculated in Sec.~\ref{sect:appendix_noise} based on the parameters summarized in Table~\ref{tab:noise_parameters}, we obtain the yaw angle noise at 1~Hz as 5.3$\times 10^{-19}$~rad~${\rm Hz}^{-1/2}$. The beam-position mismatch ratio was chosen to be $\alpha$=0.3\%. The translational displacement noise and pitch angle noise are derived by similar calculations along Eqs.~(\ref{eq:net_force})(\ref{eq:trans})(\ref{eq:net_torque_pitch})(\ref{eq:pitch}). The noise at 1~Hz is 3.9$\times 10^{-19}$~m~${\rm Hz}^{-1/2}$ in translation and 1.1$\times 10^{-18}$~rad~${\rm Hz}^{-1/2}$ in pitch. Table~\ref{tab:efficiency} summarizes the noise in each degree of freedom.

\begin{figure}[tp]
\begin{center}
\begin{tabular}{c}
\includegraphics[width=9.0cm]{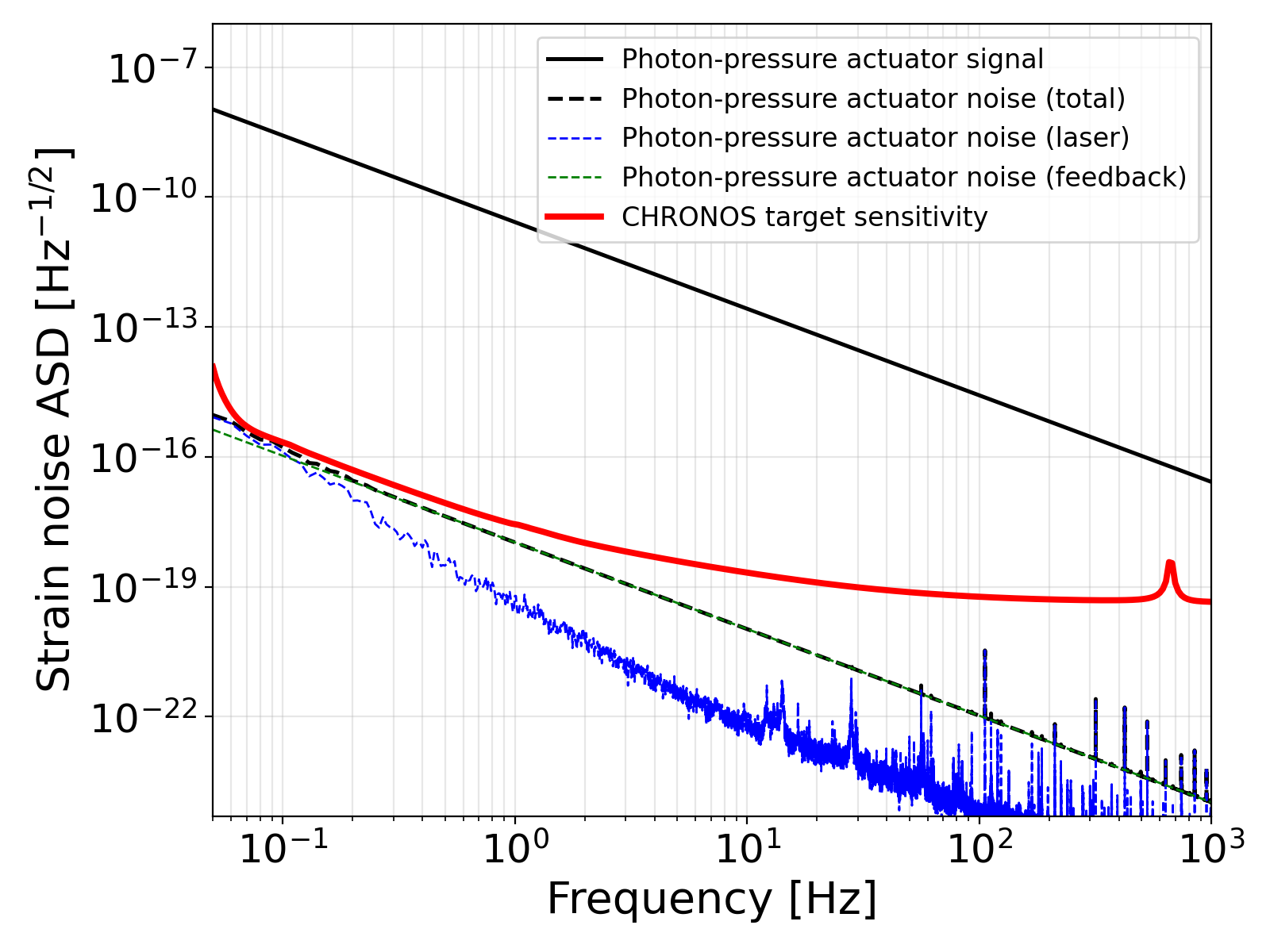}
\end{tabular}
\end{center}
\caption[]
{ \label{fig:hpcal}
Comparison of the strain-equivalent yaw rotation amplitude generated by photon-pressure actuator and the total noise budget of CHRONOS. Beam power injected to each AOM is 2.5~W. The "feedback" noise curve is a quadrature sum of offset noise, excitation noise, and shot noise.
}
\end{figure}

Figure~\ref{fig:hpcal} shows the strain-equivalent yaw signal and noise in comparison with the sensitivity curve of CHRONOS. At the frequencies sufficiently higher than the yaw resonant frequency, the strain-equivalent signal is simply derived by twice of the yaw rotation angle. The actuation noise in yaw is smaller than the designed sensitivity above 1~Hz.

\subsection{Systematic error}
\label{sect:systematic}

In this study, we consider the case where we use the photon-pressure actuator as PCal. The systematic error of angle induced by the calibrator propagates to the error of interferometer response, thus that of GW. Uncertainty of beam power, incident angles, beam positions, and moment of inertia are taken into account as sources of systematic error. 

The errors propagate as following,
\begin{equation}
\Delta \varphi_Y=\displaystyle \sum_i^4\left| \frac{\partial \varphi_Y}{\partial P_{\rm i}} \Delta P_{\rm i}\right|+\sum_i^4\left| \frac{\partial \varphi_Y}{\partial a_{\rm i}} \Delta a_{\rm i}\right|+\sum_i^4\left| \frac{\partial \varphi_Y}{\partial \phi_{\rm i}} \Delta \phi_{\rm i}\right|+\sum_i^4\left| \frac{\partial \varphi_Y}{\partial \theta_{\rm i}} \Delta \theta_{\rm i}\right|+\left|\frac{\partial \varphi_Y}{\partial I_Y}\Delta I_Y \right|.
\end{equation}

We assumed 1.25~W as the maximum power amplitude of each beam injected to the ETM. Accuracy of power relies on calibration using integrating spheres, TxPDs, and RxPD. Using the same power sensor system as our design, KAGRA assigned 0.614\% as the error of the total PCal power during its 4th joint observation~\cite{kagra_pcal_o4}. 

We evaluated the uncertainty of the beam incident angles $\theta_{\rm i}$ and $\phi_{\rm i}$ as 0.07$^\circ$ by assuming 1~mm error of the beam position at the vacuum window and 1~cm error of the position of the ETM. Beam position error at the bar surface was also assumed to be 1~mm. Error of moment of inertia was estimated to be 0.04~kg$\cdot$m$^2$ around yaw axis and 0.01~kg$\cdot$m$^2$ around pitch axis by assuming 10~g error of the mass and 1~mm error of the bar dimension.

Calculating the error propagation of these uncertainties, we estimated the systematic error of the PCal as 2.4$\times$10$^{-13}$~rad at 1~Hz, which corresponds to 1.14\% of the mean value. Uncertainties of the overall parameters and the total error are summarized in Table~\ref{tab:systematics}.

\begin{table}[tbp]
\begin{center}\scriptsize
\begin{tabular}{c|c|c|c|c}
\hline
Parameter & Main value (Unit) & Error Unit) & Relative error (\%) & Contribution to total (\%) \\
\hline\hline
$P_{\rm i}$ (${\rm i}=1,3$) & 1.25 (W) & - & 0.61 & 0.30 \\
$P_{\rm i}$ (${\rm i}=2,4$) & 1.25 (W) & - & 0.61 & 0.31 \\
$|\cos\theta_{\rm i}|$ & 1.0 & 0.07 & 0.006 & 0.006 \\
$|\cos\phi_{\rm i}|$ & 1.0 & 0.07 & 7.5$\times$10$^{-5}$ & 7.5$\times$10$^{-5}$ \\
$|r_{\rm i}^x|$ (${\rm i}=1,3$) & 0.306 (m) & 0.001 (m) & 0.33 & 0.16 \\
$|r_{\rm i}^x|$ (${\rm i}=2,4$) & 0.317 (m) & 0.001 (m) & 0.32 & 0.16 \\
$I_{\rm Y}$ & 19.9 (kg$\cdot$m$^2$) & 0.04 (kg$\cdot$m$^2$) & 0.20 & 0.20\\
\hline
$\varphi_Y$ at 1~Hz & \begin{tabular}{c} 1.3$\times$10$^{-11}$\\(rad) \end{tabular} & \begin{tabular}{c} 1.5$\times$10$^{-13}$\\(rad) \end{tabular} & 1.14 & - \\
\hline
\end{tabular}
\end{center}
\caption{Systematic uncertainty of yaw rotation angle induced by photon-pressure actuator and the relevant parameters. Contributions of power, angle, and position to the total error show the sum of beams.}
\label{tab:systematics}
\end{table}

\section{Discussion}
\label{sect:discussion}

The predicted actuation range and efficiency is sufficient for maintaining lock, although it is thousand times smaller than the electromagnetic actuator that is used for ETM of KAGRA~\cite{kagra_actuation}. The maximum force of the photon-pressure actuator is determined by laser power. When a laser at higher power is adopted as a source, we have to attenuate the power injected to the PDs to avoid saturation by inserting beam splitters or neutral density filters after the beam samplers. As discussed below, tolerance and linearity of the AOM have to be considered as well.

In the estimation of available actuation power, we simplified the AOM model to by assuming that it can be linearly swept over all range of transmittance from 0 to 1. However, the AOM can have a limitation of acceptable RF power while the AOM driver has the upper limit of the output. This constraint becomes more serious in longer wavelength because the required RF power to achieve the same diffraction efficiency is proportional to the wavelength. The AOM driver we assumed in this study can output up to 15~W, and the AOM accepts only 10~W at maximum~\cite{AOM_1550nm_M1377,AOM_1550nm_driver_RFJ080}. This leads to the maximum transmittance of 0.68 and the linear region of 4.4-7.8~V with $\pm$5\% error of the slope $dT^{(1)}/dV_{\rm AOM}$. 

The non-linearity of AOM response induces higher harmonics of injection line. It has been reported that higher harmonics was induced due to non-linearity of PCal developed in KAGRA~\cite{kagra_pcal}. Since the transfer function given by Eq.~(\ref{eq:yaw}) is inverse square of frequency, we approximate $\varphi_{\rm Y}(nf)\sim \varphi_{\rm Y}(f)/n^2$. When we write the signal-to-ratio (SNR) of rotation at the injection frequency as $G_Y$ and that of $n$-th harmonics frequency as $G_{Y}^{(n)}$, the requirement to the higher harmonics of output power is written as
\begin{equation}
\label{eq:harmonics}
\frac{P(nf)}{P(f)}=\frac{T_{\rm Y}(nf)}{T_{\rm Y}(f)}\le \frac{n^2}{G_Y G_Y^{(n)}}\frac{h(nf)}{h(f)}.
\end{equation}
The maximum yaw rotational angle calculated in Sec.~\ref{sect:actuation_efficiency} gives $G_Y>10^6$ in 0.1-10~Hz as shown in Fig.~\ref{fig:hpcal}. Inoue {\em et~al.} (2023) set the requirement of $G_Y^{(n)}\le 0.1$ and reported $P(nf)/P(f)\sim -70$~dB in the frequency range of 20-750~Hz~\cite{kagra_pcal}. Substituting CHRONOS sensitivity of $h(1~{\rm Hz})\sim 3\times 10^{-18}~{\rm Hz}^{-1/2}$, $h(2~{\rm Hz})\sim 1\times 10^{-18}~{\rm Hz}^{-1/2}$, and $h(3~{\rm Hz})\sim 7\times 10^{-19}~{\rm Hz}^{-1/2}$ into Eq.~(\ref{eq:harmonics}), the requirements to 2$f$ and 3$f$ harmonics are derived as $1.2\times 10^{-6}$ ($-$118~dB) and $3.9\times 10^{-6}$ ($-$108~dB)~\cite{CHRONOS}. To achieve these requirements to higher harmonics, the operation range of AOM should be constrained in the linear region.

Casting an eye on another product, MT80-A0.7-1300.1600 AOM and MODA80-B41k51k-344575 driver provided by AA Opto-Electronic can fully use the transmittance range reaching to above 0.75~\cite{AOM_1550nm_MT80,AOM_1550nm_driver_MODA80}. On the other hand, it tolerates only 10~W/mm${}^2$ which is smaller than our beam power assuming beam diameter of 0.3~mm at the AOM. Developing an AOM for high power can be essential to increase actuation range of the photon-pressure actuator.

If the actuation force is also sufficient for lock acquisition, the photon-pressure actuator can replace the recoil mass. The requirement to the actuation force $F_{\rm req}$ for lock acquisition is determined by the velocity of the ETM~\cite{kagra_actuation}. In the case of rotation,
\begin{equation}
\label{eq:velocity}
F_{\rm req}\simeq I_{\rm Y}\frac{\omega_{\rm vel}^2}{d},
\end{equation}
where $\omega_{\rm vel}$ is angular velocity and $d$ is the cavity linewidth in the unit of length. We approximate $d$ with a half of the laser wavelength, namely 775~nm. Given that the strain-equivalent seismic noise of CHRONOS is estimated to be 1$\times 10^{-12}~{\rm Hz}^{-1/2}$ at 0.05~Hz, the angular velocity becomes 7.9$\times 10^{-14}~{\rm rad/s}$ for a single bar~\cite{CHRONOS}. Thus, Eq.~(\ref{eq:velocity}) gives $F_{\rm req}\simeq 1.6\times 10^{-19}$~N. The predicted actuation force of the photon-pressure actuator satisfies this, but the measurement of seismic noise on site is necessary for assessment of the actuation systems.

We showed that the dominant noise source of the photon-pressure actuator is the DAC noise induced with the excitation through the feedback circuit. In order to achieve sufficiently low noise by dewhitening, developing a DAC having larger output range is necessary. 

Once we suppress the DAC noise at the sufficient level, the laser intensity noise becomes dominant. Increasing the OFS gain and transimpedance gain can improve the feedback gain of the power stabilization circuit. Possible solution for further stabilization could be temperature control of the laser diode or two-staged power stabilization. 

In order to reduce systematic error, the error of power $P_{\rm i}$ is the largest source and the first to be considered. The combination of TxPD and the integrating sphere which is used for power calibration is called Gold Standard and annually calibrated among LIGO, KAGRA, and NIST~\cite{ligo_pcal,kagra_pcal}. If the uncertainty of $P_{\rm i}$ reduces to a half, namely 0.31\%, the total systematic error reduces to 0.84\%. Additionally, when the error of moment of inertia $I_{\rm Y}$ can reduce to one-tenth by measuring the geometry of ETM after fabrication, the total systematic error further reduces to 0.66\%.

Although another calibration device named gravity field calibrator (GCal) predicts a better accuracy of 0.24\% in 0.1-10~Hz when it is configured for CHRONOS, the photon-pressure actuator as PCal is still preferred for calibration in higher frequencies because of the limitation of rotation speed of GCal~\cite{chronos_gcal}. Combination of the photon-pressure actuator and GCal is expected to further improve the calibration accuracy in a wide frequency range. It has been predicted to achieve 0.17\% accuracy in the configuration of KAGRA by calibrating the laser power of the PCal using the detector response generated by GCal~\cite{kagra_gcal_pcal}. 

In this study, we did not consider the detailed parameters in the receiver module because it does not contribute to the actuation force and noise. However, it is important to monitor the beam position by QPDs because the beam position error is a source of the systematic error. Furthermore, the beam position offset can excite the bulk deformation mode that causes surface displacement, thus inducing an excess systematic error.
\section{Conclusion}
\label{sect:conclusion}

We proposed an actuation principle and signal model for photon-pressure actuator applied to a torsion-bar-based GW detector. Its core technique is a power control circuit using AOM under feedback by OFS. We designed the transmitter module, periscope, and receiver module for 1550~nm wavelength laser. The actuation force and torque were simply modeled by laser power, incident angle, and mass property of the torsion bar. We chose CHRONOS experiment as a model platform and determined the incident points to minimize bulk deformation of the ETM based on FEA analysis. Assuming four beams of 2.5~W power, we evaluated the maximum torque amplitude and actuation efficiency in yaw rotation to be $1.0\times 10^{-8}$~N$\cdot$m and $6.6\times 10^{-13}$~rad/V, which are sufficient for both lock acquisition and maintaining lock. The estimated strain-equivalent noise is 5.3$\times 10^{-19}$~rad/${\rm Hz}^{-1/2}$ which is sufficiently lower compared with CHRONOS sensitivity at 1~Hz. The systematic error when it is used as PCal was evaluated to be 1.14\%. 

The photon pressure actuator, which can be designed as a two-in-one integrated system with Photon Calibrator, is free from source vibration and environmental magnetic fields, and therefore can mitigate low-frequency actuation noise. It is beneficial for GW experiments which aims to detect heavier binary black hole mergers. It potentially simplifies suspension structure, feedback operation, and calibration model. The photon-pressure actuator can be a powerful tool for the next-generation GW experiments in both scientific and operational aspects.

\acknowledgments     
We thank Masashi Hazumi for his academic advice during the preparation of this manuscript. We appreciate Rick Savage and Sadakazu Haino for discussion on the evaluation method of bulk deformation. Chao Shiuh helped our mechanical simulation. We are grateful to Tsung-Chieh Ho, Ko-Han Chen, Aloysius Niko, and Cheng-Han Chan, who contributed to develop the input optics of CHRONOS. Avani Patel, Afif Ismail, Hsiang-Chieh Hsu, and Henry Tsz-King Wong provided fruitful discussions on the sensitivity calculation of CHRONOS. Y.I. and D.T. acknowledges support from NSTC, CHiP, and Academia Sinica in Taiwan under Grant No.114-2112-M-008-006- and No.AS-TP-112-M01.

\appendix
\appendix
\section{Noise model}
\label{sect:appendix}

\subsection{Formalism}
\label{sect:app_formalism}

\begin{figure}[tp]
\begin{center}
\begin{tabular}{c}
\includegraphics[width=8.0cm]{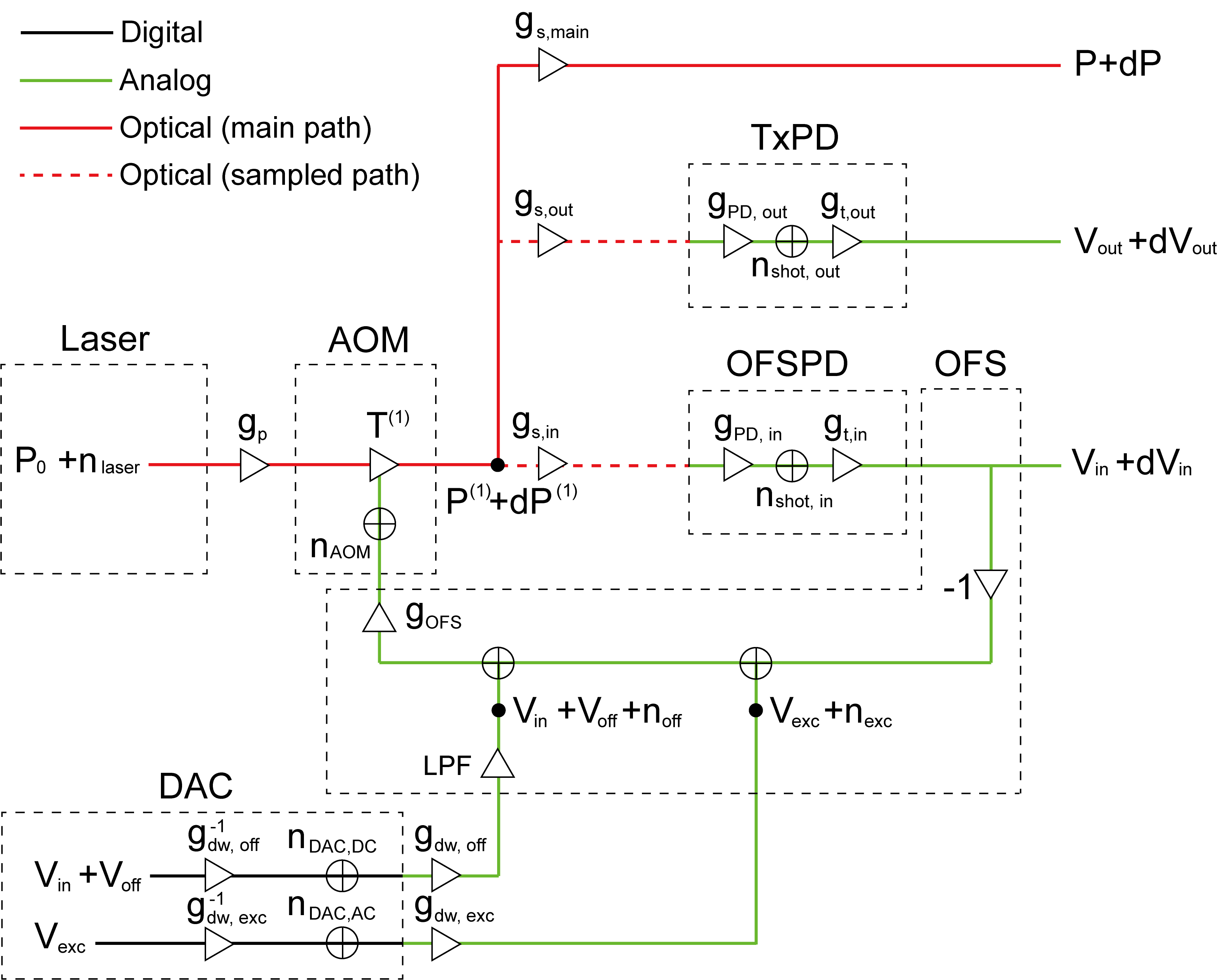}
\end{tabular}
\end{center}
\caption[]
{ \label{fig:feedback_diagram}
Schematic diagram of the power stabilization circuit adopting OFS. The output voltages of OFSPD, TxPD, and the output power to the ETM are shown.
}
\end{figure}

We modeled the relative power noise of each beam based on the feedback control diagram shown in Fig.~\ref{fig:feedback_diagram}. The relevant  parameters are summarized in Table~\ref{tab:noise_parameters}.

Each beam right after the beam splitter has DC power $P_0$ and absolute power noise $n_{\rm laser}$. Energy loss on the path from the laser to AOM is expressed by an optical efficiency $g_{\rm p}$. The AOM transmittance of the first-order diffraction beam is $T^{(1)}$, but the feedback signal $n_{\rm sum}$ slightly corrects it in proportional to $dT^{(1)}/dV_{\rm AOM}$, the slope of transmittance to the analog input voltage. The DC term and perturbation of the first-order beam power right after the AOM can be expressed as
\begin{equation}
\label{eq:V_tot_P}
P^{(1)}+dP^{(1)}=\left( P_0+n_{\rm laser} \right)g_{\rm p}\left( T^{(1)}+\frac{dT^{(1)}}{dV_{\rm AOM}}n_{\rm sum} \right),
\end{equation}
where
\begin{equation}
\label{eq:P1}
P^{(1)}\equiv P_0 g_{\rm p} T^{(1)}.
\end{equation}

The first-order beam is split into three by a diffractive beam sampler of sampling ratio of $g_{\rm s,in}$ and $g_{\rm s,out}$. The energy fraction of the main beam is 
\begin{equation}
g_{\rm s,main}\equiv 1-g_{\rm s,in}-g_{\rm s,out}.
\end{equation}
Another two beams are detected by the in-loop and out-loop PDs, which are labeled as OFSPD and TxPD. The power is converted to current by quantum efficiency $g_{\rm PD}$, gains shot noise $n_{\rm shot}$, and then converted to voltage by transimpedance gain $g_{\rm t}$. The DC voltage detected by the in-loop OFSPD is
\begin{equation}
V_{\rm in}=P^{(1)} g_{\rm s,in}g_{\rm PD,in}g_{\rm t,in}.
\end{equation}

The in-loop photodetector is connected to the OFS which applies negative feedback to the sensed signal. The OFS also applies offset $V_{\rm in}+V_{\rm off}$ and excitation $V_{\rm exc}$ provided from the DAC. Each of the offset and excitation has noise originated to DAC, $n_{\rm off}$ and $n_{\rm exc}$. After the amplifier gain $g_{\rm OFS}$ and AOM driver noise $n_{\rm AOM}$, the total feedback signal applied to the AOM is 
\begin{equation}
\label{eq:nsum_P}
\begin{array}{ll}
n_{\rm sum}=&\left[ -\left\{\left( P^{(1)}+dP^{(1)} \right)g_{\rm s,in}g_{\rm PD,in}+n_{\rm shot,in}\right\}g_{\rm t,in}+V_{\rm in}+n_{\rm off}+V_{\rm exc} +n_{\rm exc} \right] g_{\rm OFS}+n_{\rm AOM}\\
\phantom{n_{\rm sum}}=&\left\{ -\left(dP^{(1)}g_{\rm s,in}g_{\rm PD,in}+n_{\rm shot,in}\right)g_{\rm t,in}+n_{\rm off}+V_{\rm exc} +n_{\rm exc} \right\} g_{\rm OFS}+n_{\rm AOM}
\end{array}
\end{equation}
Note that the $V_{\rm off}$ term is used for keeping the AOM at the operating point $T^{(1)}$ and does not appear as an excess signal.

We solve Eqs.~(\ref{eq:V_tot_P})(\ref{eq:nsum_P}) for $dP^{(1)}$. It can be expressed by a sum of excitation and noise terms as
\begin{equation}
\label{eq:dV_P}
dP^{(1)}=dP_{\rm exc}^{(1)}+\delta P^{(1)}
\end{equation}
where
\begin{equation}
\label{eq:dV_exc_P}
dP_{\rm exc}^{(1)}\equiv\frac{1}{1+G} P_0 g_{\rm p}\frac{dT^{(1)}}{dV_{\rm AOM}}g_{\rm OFS} V_{\rm exc}, 
\end{equation}
\begin{equation}
\label{eq:delta_V_P}
\begin{array}{ll}
\delta P^{(1)}&=\frac{1}{1+G}\\
& \cdot \left[ g_{\rm p}T^{(1)} n_{\rm laser}\right.\\
& -P_0 g_{\rm p}\frac{dT^{(1)}}{dV_{\rm AOM}}g_{\rm t,in}g_{\rm OFS}n_{\rm shot,in}\\
& +P_0 g_{\rm p}\frac{dT^{(1)}}{dV_{\rm AOM}}g_{\rm OFS}\left( n_{\rm off}+n_{\rm exc} \right)\\
& \left.+P_0 g_{\rm p}\frac{dT^{(1)}}{dV_{\rm AOM}}n_{\rm AOM} \right].
\end{array}
\end{equation}
Here we defined the open-loop gain
\begin{equation}
\label{eq:G}
G\equiv P_0 g_{\rm p}\frac{dT^{(1)}}{dV_{\rm AOM}}g_{\rm s,in}g_{\rm PD,in}g_{\rm t,in}g_{\rm OFS}.
\end{equation}
Output from each point in the feedback circuit is a product of the open-loop transfer function $1/(1+G)$, additive inputs including noise, and gains on the path to the output.

The power injected to the ETM is
\begin{equation}
\label{eq:dP}
P=g_{\rm s,main}P^{(1)}, \ dP_{\rm exc}=g_{\rm s,main}dP_{\rm exc}^{(1)}, \ \delta P=g_{\rm s,main} \delta P^{(1)}.
\end{equation}
We define the RPN of this transmission power by $\delta P/P_0$ for potential discussion of increasing the power of laser source.

Considering that the laser intensity noise is in-phase among the four beams while the others are independent, we separately define the laser intensity component $\delta P_{\rm laser}$ and the feedback-induced component $\delta P_{\rm feedback}$ in the power noise injected to the ETM.
\begin{equation}
\label{eq:delta_P_laser}
\delta P_{\rm laser}\equiv \frac{1}{1+G}g_{\rm p}T^{(1)}g_{\rm s,main}n_{\rm laser},
\end{equation}
\begin{equation}
\label{eq:delta_P_feedback}
\delta P_{\rm feedback}\equiv \frac{1}{1+G}P_0 g_{\rm p} \frac{dT^{(1)}}{dV_{\rm AOM}}g_{\rm s,main}\sqrt{g_{\rm t,in}^2 g_{\rm OFS}^2 n_{\rm shot,in}^2+g_{\rm OFS}^2 \left( n_{\rm off}^2+n_{\rm exc}^2 \right)+n_{\rm AOM}^2}.
\end{equation}

With similar calculations, the output voltage of the out-loop OFSPD is derived as
\begin{equation}
\label{eq:dV_exc_out}
dV_{\rm exc,out}\equiv \frac{1}{1+G}P_0 g_{\rm p}\frac{dT^{(1)}}{dV_{\rm AOM}}g_{\rm OFS}g_{\rm s,out}g_{\rm PD,out}g_{\rm t,out} V_{\rm exc},
\end{equation}
\begin{equation}
\label{eq:delta_V_out}
\begin{array}{ll}
&\delta V_{\rm out}=\frac{1}{1+G}\\
& \cdot \left[ g_{\rm p}T^{(1)}g_{\rm s,out}g_{\rm PD,out}g_{\rm t,out} n_{\rm laser} \right.\\
& -P_0 g_{\rm p}\frac{dT^{(1)}}{dV_{\rm AOM}}g_{\rm t,in}g_{\rm OFS}g_{\rm s,out}g_{\rm PD,out}g_{\rm t,out}n_{\rm shot,in} \\
& +P_0 g_{\rm p}\frac{dT^{(1)}}{dV_{\rm AOM}}g_{\rm OFS}g_{\rm s,out}g_{\rm PD,out}g_{\rm t,out}\left( n_{\rm off}+n_{\rm exc} \right) \\
& +P_0 g_{\rm p}\frac{dT^{(1)}}{dV_{\rm AOM}}g_{\rm s,out}g_{\rm PD,out}g_{\rm t,out}n_{\rm AOM} \\
& \left. +g_{\rm t,out}n_{\rm shot,out}\right].
\end{array}
\end{equation}

The output voltage of the in-loop OFSPD is
\begin{equation}
\label{eq:dV_exc_in}
dV_{\rm exc,in}\equiv\frac{1}{1+G}P_0 g_{\rm p}\frac{dT^{(1)}}{dV_{\rm AOM}}g_{\rm OFS}g_{\rm s,in}g_{\rm PD,in}g_{\rm t,in} V_{\rm exc},
\end{equation}
\begin{equation}
\label{eq:delta_V_in}
\begin{array}{ll}
&\delta V_{\rm in}=\frac{1}{1+G}\\
& \cdot \left[ g_{\rm p}T^{(1)}g_{\rm s,in}g_{\rm PD,in}g_{\rm t,in} n_{\rm laser} \right.\\
& +g_{\rm t,in}n_{\rm shot,in} \\
& +P_0 g_{\rm p}\frac{dT^{(1)}}{dV_{\rm AOM}}g_{\rm OFS}g_{\rm s,in}g_{\rm PD,in}g_{\rm t,in}\left( n_{\rm off}+n_{\rm exc} \right) \\
& \left. +P_0 g_{\rm p}\frac{dT^{(1)}}{dV_{\rm AOM}}g_{\rm s,in}g_{\rm PD,in}g_{\rm t,in}n_{\rm AOM} \right].
\end{array}
\end{equation}

\begin{table}[tbp]
\begin{center}\scriptsize
\begin{tabular}{c|c|c}
\hline
Description & Symbol & \begin{tabular}{c}Mean value\\Unit)\end{tabular}\\
\hline\hline
Initial power of each beam & $P_0$ & 2.5 (W) \\
Transmittance of 1st order beam at AOM & $T^{(1)}$ & 0.5 \\
Slope of 1st order beam transmission & $dT^{(1)}/dV_{\rm AOM}$ & 0.12 (V$^{-1}$) \\
DAC voltage for OFS offset & $V_{\rm off}$ & 9.98 (V) \\
DAC voltage amplitude for OFS excitation & $V_{\rm exc}$ & 9.98 (V) \\
Optical efficiency before AOM & $g_{\rm p}$ & 1 \\
Energy sampling ratio of in-loop path & $g_{\rm s,in}$ & 0.001 \\
Energy sampling ratio of out-loop path & $g_{\rm s,out}$ & 0.001 \\
Quantm efficiency of in-loop OFSPD & $g_{\rm PD,in}$ & 0.95 (A/W) \\
Quantum efficiency of out-loop OFSPD & $g_{\rm PD,out}$ & 0.95 (A/W) \\
Transimpedance gain of in-loop OFSPD & $g_{\rm t,in}$ & 5300 (V/A) \\
Transimpedance gain of out-loop OFSPD & $g_{\rm t,out}$ & 5300 (V/A) \\
OFS gain & $g_{\rm OFS}$ & 67 \\
\hline
\end{tabular}
\end{center}
\caption{Parameters in the feedback circuit for power control.}
\label{tab:noise_parameters}
\end{table}

\subsection{Assumed parameters}
\label{sect:appendix_parameters}

\subsubsection{Laser source power $P_0$}
We split a 10~W beam from the laser source FC-1550-10W, provided by CNI, into four~\cite{Laser_FC-1550-10W}. Neglecting the loss and imbalance of the beam splitters, the $P_0$ of each beam becomes to 2.5~W.

\subsubsection{Optical efficiency before AOM $g_{\rm p}$}
The major loss source before AOM in our transmittance module is the polarized beam splitters. We assumed that the polarization of the beam can be controlled by the half-wave plate. Hence, we approximated $g_{\rm p}$ to 1.

\subsubsection{AOM transmittance $T^{(1)}$ and its slope $dT^{(1)}/dV_{\rm AOM}$}
We assumed a combination of M1377-aQ80L-1 (1.5um) AOM and RFJ080-1-15 driver with 24~V power supply~\cite{AOM_1550nm_M1377}\cite{AOM_1550nm_driver_RFJ080}. The peak RF power required for the maximum diffraction efficiency is $P_{\rm max}=$26~W. On the other hand, the AOM driver outputs 15~W at the analog input of $V_{\rm max}=$10~V. We modeled the AOM transmittance to the RF power $P_{\rm AOM}$ and corresponding driving voltage $V_{\rm AOM}$ by
\begin{equation}
T^{(1)\prime}=T_{\rm 1,max}\sin^2 \frac{2\pi}{4V_{\rm max}}V_{\rm AOM},
\end{equation}
where
\begin{equation}
V_{\rm AOM}=\sqrt{\frac{10^2 P_{\rm AOM}}{15}}.
\end{equation}

\begin{figure}[tp]
\begin{center}
\begin{tabular}{c}
\includegraphics[width=8.0cm]{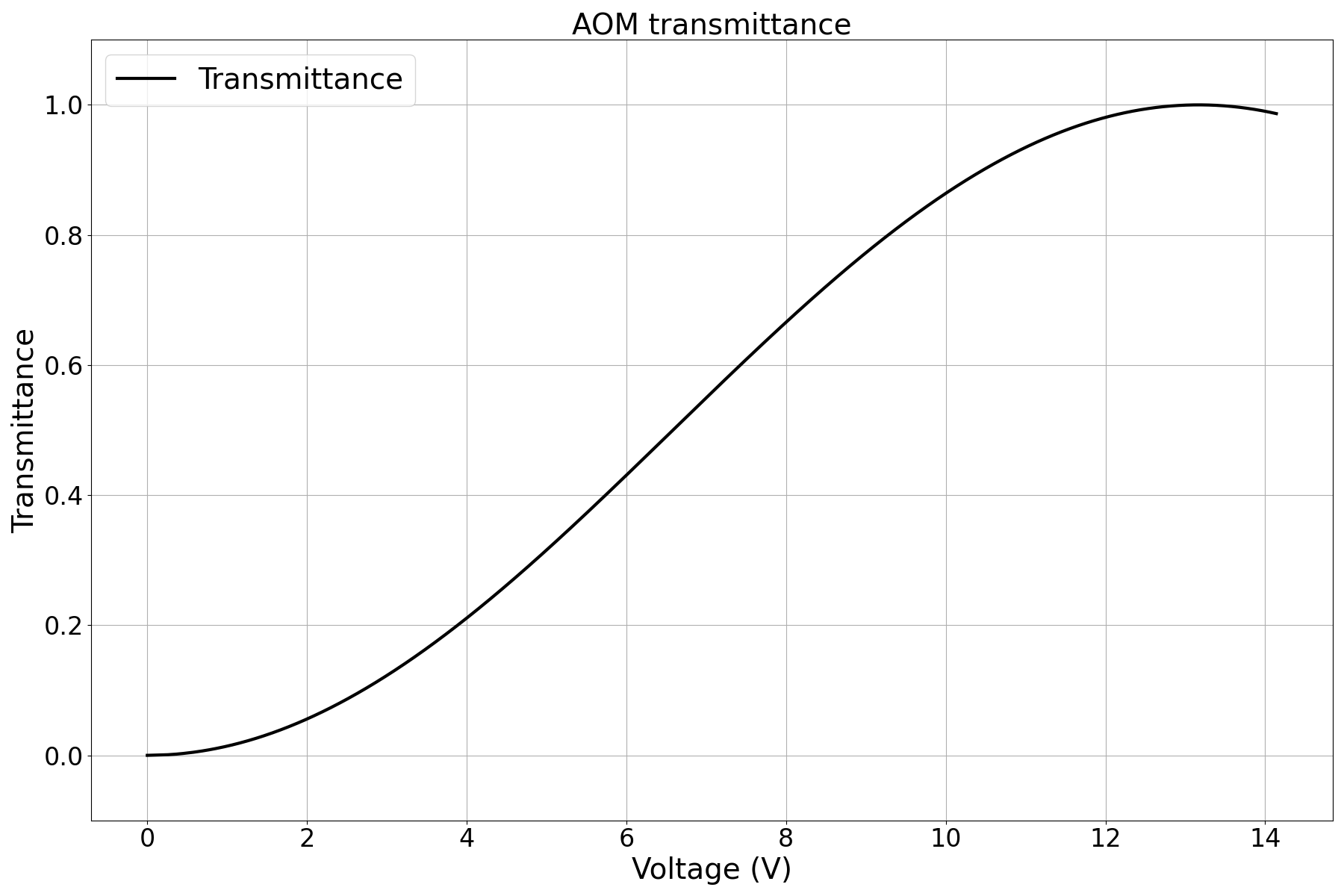}
\end{tabular}
\end{center}
\caption[]
{ \label{fig:AOM_nominal}
Modeled transmittance curve of M1377-aQ80L-1 (1.5um) AOM and RFJ080-1-15 driver with 24~V power supply.
}
\end{figure}

Figure~\ref{fig:AOM_nominal} shows the modeled transmittance curve. To fully use the input laser power for excitation, we set the operation point at $T^{(1)}$=0.5, where $V_{\rm AOM}=$6.58~V and the slope $dT^{(1)}/dV_{\rm AOM}$ is 0.12~V$^{-1}$.

\subsubsection{Energy sampling ratio $g_{\rm s}$}
Power injected to each PD should be attenuated to the order of milliwatt to avoid saturation. We assumed SA-222-G-Y-A diffractive beam sampler provided by HOLO/OR whose the energy sampling ratio is nominally 0.1\%~\cite{Sampler_1550_0p1}. The sampled power of each beam becomes 1.25~mW which we accept.

Substituting $g_{\rm p}=$1, $T^{(1)}$=0.5, and $g_{\rm s,main}=$0.998 into Eqs.~(\ref{eq:P1})(\ref{eq:dP}), we obtain $P\sim$1.25~W as an DC actuation power. Sweeping $T^{(1)}$ from 0 to 1 by excitation signal, the actuation amplitude is 1.25~W at most.

\subsubsection{OFS offset $V_{\rm off}$}
The $V_{\rm off}$ provides a bias to keep the AOM at the operation point. The relationship between $V_{\rm off}$ and $V_{\rm AOM}$ follows
\begin{equation}
\label{eq:aom_offset}
V_{\rm AOM}=\frac{1}{1+G}g_{\rm OFS}V_{\rm off}.
\end{equation}
Since the coefficient $g_{\rm OFS/}(1+G)$ is not more than 1, it requires $V_{\rm off}$ to be at the order of 1~V to realize $V_{\rm AOM}=$6.58~V. 

However, the DAC channel also has to output $V_{\rm in}$ to cancel the DC offset of the detected power. The General Standard PCIe-16AO16-16-F0-DF DAC used in LIGO, which divides $\pm$10 V range by 16 bits, individual channel has not large margin to output $V_{\rm off}+V_{\rm in}$~\cite{general_standard_dac}. Since $V_{\rm in}$ is controlled by the transimpedance gain $g_{\rm t,in}$, the constraint to $V_{\rm in}$ limits the feedback gain $G$ as well as noise suppression. 

Therefore, we require a DAC that can output $\pm$20~V by a single channel. Under this assumption, later we chose a suitable value for $g_{\rm t,in}$ to satisfy $V_{\rm off}+V_{\rm in}\le$20~V and $V_{\rm AOM}=$6.58~V simultaneously.

\subsubsection{OFS excitation $V_{\rm exc}$}
This term directly determines amplitude of the actuation power. According to our AOM transmittance curve, the possible maximum amplitude of $V_{\rm AOM}$ is 6.58~V. Since the $V_{\rm exc}$ is converted to $V_{\rm AOM}$ by the same coefficient as Eq.~(\ref{eq:aom_offset}), we obtain the amplitude of $V_{\rm exc}$ as 9.98~V with the value of $g_{\rm t,in}$ assumed below.

Besides, we can consider dewhitening method that attenuates the signal and noise before OFS to suppress DAC noise. To realize noise reduction, we consider to output 20~V from the DAC and attenuate it to 9.98~V. 

\subsubsection{Quantum efficiency of OFSPD $g_{\rm PD}$}
We assumed C30665GH InGaAs photodetector provided by Excelitas for both in-loop and out-loop PDs~\cite{Excelitas_PD}. Its quantum efficiency at 1550~nm is nominally 0.95~A/W. 

\subsubsection{Transimpedance gain of PD $g_{\rm t}$}
The transimpedance gain can be flexibly adjusted by a gain resistor on the readout circuit~\cite{ligo_PD}. It is the dominant term in the total feedback gain $G$. We chose $g_{\rm t,in}=g_{\rm t,out}=$5300 to realize $V_{\rm in}$=6.29~V and $V_{\rm off}=$9.98~V when the DC offset of $V_{\rm AOM}$ is 6.58~V.

\subsubsection{OFS gain $g_{\rm OFS}$}
The $g_{\rm OFS}$ should be adjusted according to the targeted $G$. It mainly contributes to suppression of the $n_{\rm laser}$ term. We assumed $g_{\rm OFS}=$67 to realize $G\ge$100.

\subsection{Noise budget}
\label{sect:appendix_noise}

\subsubsection{Laser intensity noise $n_{\rm laser}$}
\begin{figure}[tp]
\begin{center}
\begin{tabular}{c}
\includegraphics[width=8.0cm]{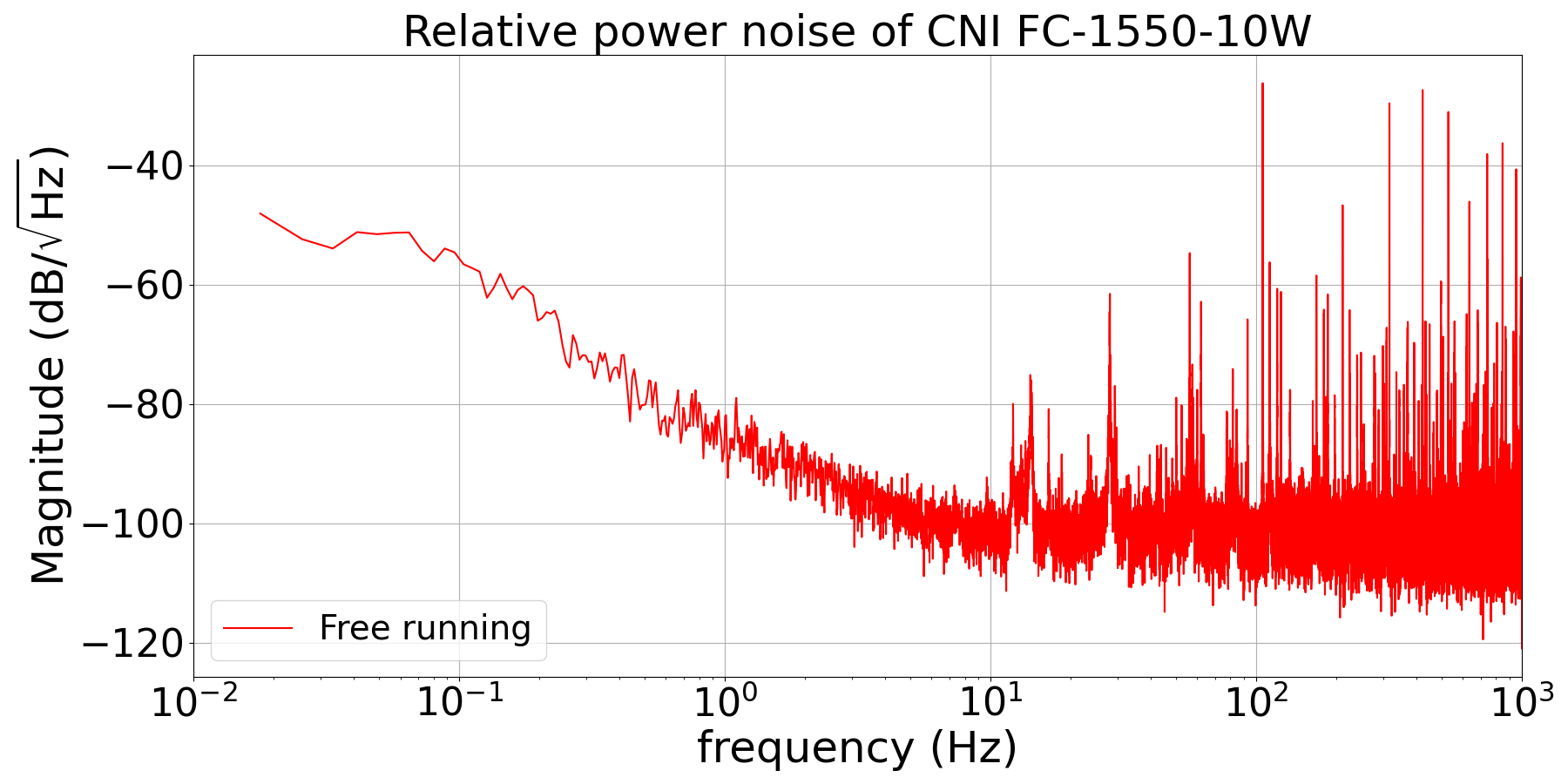}
\end{tabular}
\end{center}
\caption[]
{ \label{fig:RIN_1550nm}
Relative power noise of the FC-1550-10W laser. The DC voltage sensed by PD was 2.02 V.
}
\end{figure}

We measured the RPN of FC-1550-10W fiber laser and obtained noise floor of -100~dB/${\rm Hz}$ as shown in Fig.~\ref{fig:RIN_1550nm}. We calculated $n_{\rm laser}$ by multiplying this RPN by the design value of $P_0$.

\subsubsection{Shot noise $n_{\rm shot}$}
Each photodetector has a noise current originated to the quantum fluctuation of the photon number. It has a relationship with the sensed current as
\begin{equation}
n_{\rm shot,loop}=\sqrt{2e I_{\rm loop}},    
\end{equation}
where
\begin{equation}
I_{\rm loop}=P_0 g_{\rm p} T^{(1)} g_{\rm s,loop} g_{\rm PD,loop} \ {\rm (loop=in,out)},    
\end{equation}
at the first order. The assumed parameters in Table~\ref{tab:parameters} gives $I_{\rm in}=I_{\rm out}\sim 1.2$~mA, thus $n_{\rm shot,in}=n_{\rm shot,out}\sim 19.5$~pA/${\rm Hz}^{-1}$.

\subsubsection{Offset noise $n_{\rm off}$}
The DAC noise propagates to the OFS offset voltage. We assumed -130~dBV/${\rm Hz}$ white noise for the intrinsic DAC noise as reported in KAGRA~\cite{binhua}. It does not depend on DC voltage. 

The offset input port of OFS currently used in LIGO and KAGRA has a low-pass filter with a pole at 0.8~Hz~\cite{OFS_front}. We can further suppress the noise at the excitation frequencies without changing the DC voltage by inserting an analog dewhitening filter, denoted by $g_{\rm dw,off}$ in Fig.~\ref{fig:feedback_diagram}, between the DAC and OFS~\cite{Rana_thesis}. In  dewhitening, we output higher voltage amplified by anti-dewhitening filter $g_{\rm dw,exc}^{-1}$ from the DAC and attenuate it at the dewhitening filter $g_{\rm dw,exc}$ in the analog circuit. We conservatively assume -20~dB attenuation in 0.1-1000~Hz. Therefore, we obtain
\begin{equation}
n_{\rm off}=10^{-150/20}\frac{2\pi i f}{2\pi (i f+0.8)}.
\end{equation}

\subsubsection{Excitation noise $n_{\rm exc}$}
The excitation noise originates to the DAC noise as well as the offset noise. It is reported as -120~dBV/${\rm Hz}$ white noise when the output is AC voltage at the order of 1~V peak-to-peak magnitude~\cite{binhua}. On the contrary to the offset, it does not pass through the low-pass filter. Dewhitening by attenuation from 20~V to 9.98~V leads suppression of $-$6~dB. The noise formula is given by
\begin{equation}
n_{\rm exc}=10^{-126/20}.
\end{equation}

\subsubsection{AOM driver noise $n_{\rm AOM}$}
The circuit noise in the RF driver for AOM potentially perturbs the operation point of AOM through the fluctuation of RF power. However, we neglected it as DECIGO reported that the intensity noise did not change by AOM-driving voltage~\cite{Takahashi_fiberlaser}.  

\subsubsection{Total noise}
\begin{figure}[tp]
\begin{center}
\begin{tabular}{c}
\includegraphics[width=8.0cm]{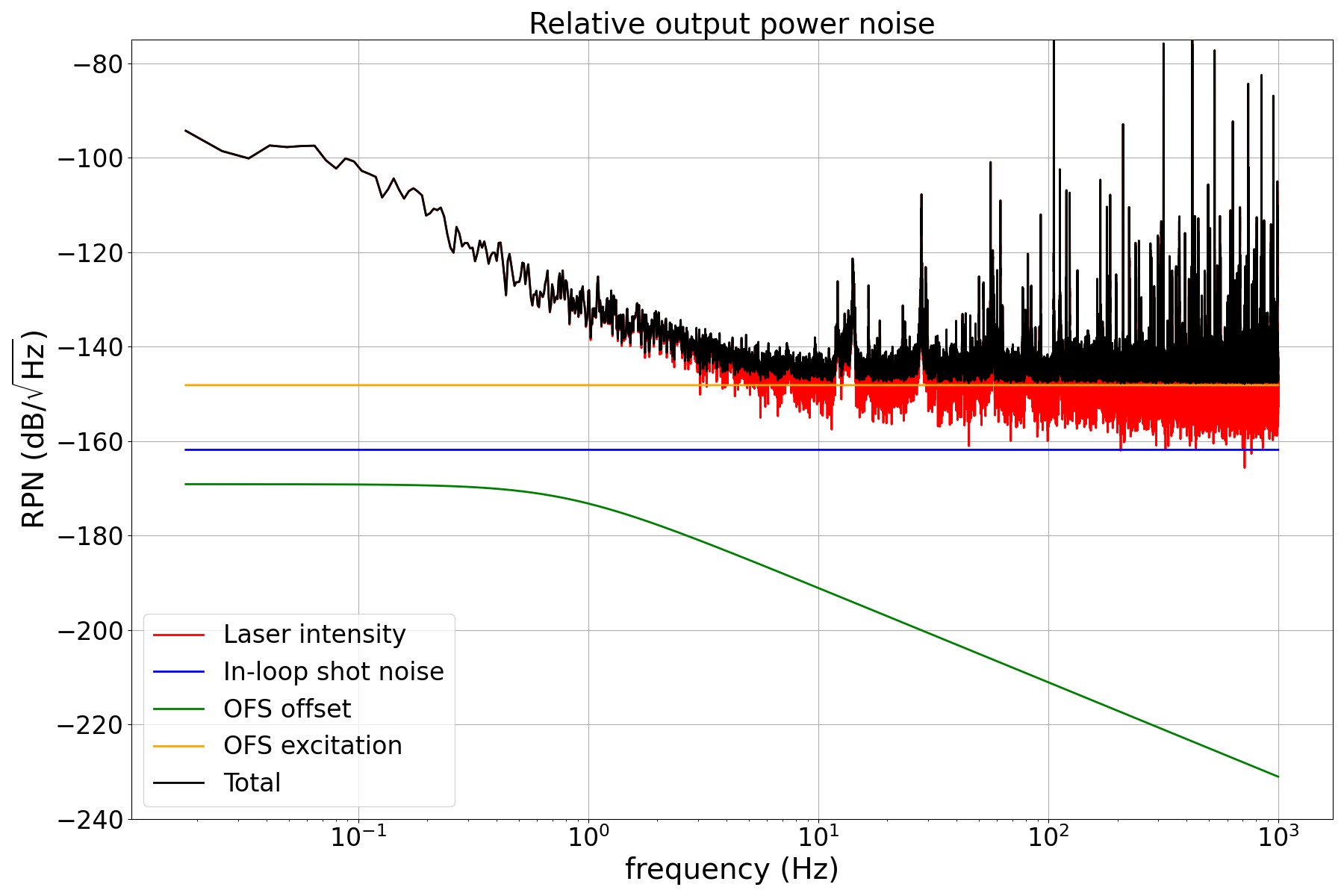}
\end{tabular}
\end{center}
\caption[]
{ \label{fig:laser_total_noise}
Relative power noise budget of the total noise power injected to the ETM. Normalized by $P_0=$2.5~W. Each noise components were multiplied by feedback coefficients defined in Eq.~(\ref{eq:delta_V_P}).
}
\end{figure}

Substituting all these noise properties and assumed parameters into Eqs.~(\ref{eq:delta_V_P})(\ref{eq:G})(\ref{eq:dP}), we obtain the total noise curve of the power injected to the ETM as shown in Fig.~\ref{fig:laser_total_noise}.

\bibliography{reference}   
\bibliographystyle{chronosbib} 

\end{spacing}
\end{document}